\documentclass[prb,aps,onecolumn,showpacs]{revtex4}
\usepackage{latexsym}
\usepackage{graphicx}
\usepackage{mathrsfs}
\usepackage{amssymb}
\usepackage{amsmath}
\usepackage{amsfonts}

\begin{document}

\title{ Schwinger-Keldysh canonical formalism for electronic Raman scattering }

\author{Yuehua Su}
\affiliation{ Department of Physics, Yantai University, Yantai 264005, P. R. China}
\email{suyh@ytu.edu.cn}

\begin{abstract}
Inelastic low-energy Raman and high-energy X-ray scatterings have made great progress
in instrumentation to investigate the strong electronic correlations in matter.
However, theoretical study of the relevant scattering spectrum is still a
challenge. In this article, we present a Schwinger-Keldysh canonical perturbation
formalism for the electronic Raman scattering, where all the resonant, non-resonant
and  mixed responses are considered uniformly. We show how to use this formalism to
evaluate the cross section of the electronic Raman scattering off an one-band superconductor.
All the two-photon scattering processes from electrons, the non-resonant charge
density response, the elastic Rayleigh scattering, the fluorescence, the intrinsic
energy-shift Raman scattering and the mixed response, are included. In the mean-field
superconducting state, Cooper pairs contribute only to the non-resonant response.
All the other responses are dominated by the single-particle excitations and are strongly
suppressed due to the opening of the superconducting gap. Our formalism for the electronic Raman
scattering can be easily extended to study the high-energy resonant inelastic X-ray scattering.

\end{abstract}


\maketitle

\section{Introduction} \label{sec1}

Inelastic low-energy Raman and high-energy X-ray scatterings have become
powerful tools to study the strong electronic correlations in matter\cite{DevereauxRMP2007,KotaniRMP2001,AmentRMP2011}.
While the instrumental technique of the light scattering is in rapid progress,
the theoretical study of the scattering spectrum is in less development.
Two main difficulties suppress the theoretical study of the light scattering off the
strongly correlated electrons. The first difficulty stems from the complexity of
the strongly correlated electrons themselves. We now have no well-defined theoretical
formalism for the various electronic correlations in such as the high-Tc cuprates,
iron-based superconductors and heavy fermions etc., where the multiple comparable
energy scales and different degrees of freedom are strongly correlated.
The other difficulty lies in the description of the inelastic light scattering processes.
Unlike the single-particle scattering technique
such as the angle-resolved photoemission spectroscopy (ARPES), neutron scattering etc.,
Raman and X-ray scatterings involve two-step photon-in photon-out processes.
The cross section of the ARPES or neutron scattering is determined by the scattering
correlation function which can be studied in perturbation formalism by the
fluctuation-dissipation theorem.
However, a simple extension of this formalism for the two-photon scattering
is fail because the fluctuation-dissipation theorem is now invalid.
We thus have no reliable perturbation formalism to study the scattering correlation function
in Raman and X-ray scatterings.

In this article, we focus our study on the second difficulty. We show that it can be
overcome by introducing the Schwinger-Keldysh contour time formalism, which has been
well established for non-equilibrium physics\cite{Schwinger,Keldysh,Jorgen}.  In this article,
we present a Schwinger-Keldysh perturbation formalism to evaluate the cross
section of the electronic Raman scattering. The formalism for the high-energy resonant inelastic
X-ray scattering (RIXS) can be established in a similar procedure.

\begin{figure}[ht]
\includegraphics[width=0.3\columnwidth]{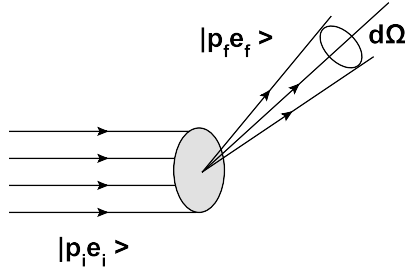}
\caption{Schematic light scattering from a target matter.
$\vert \mathbf{p}_i \mathbf{e}_i \rangle$ and $\vert \mathbf{p}_f \mathbf{e}_f \rangle$
denote the incoming photon state with momentum $\mathbf{p}_i$ and polarization
$\mathbf{e_i}$ and the scattered photon state with momentum $\mathbf{p}_f$ and polarization
$\mathbf{e_f}$, respectively. }
\label{fig1.1}
\end{figure}

Our starting point is the differential cross section of the inelastic light
scattering. Consider a two-step photon-in photon-out scattering as
shown schematically in Fig. \ref{fig1.1}. The incident photon with momentum $\mathbf{p}_i$
and polarization $\mathbf{e}_i$ is absorbed by the electrons of the target matter which
then emits photon with momentum $\mathbf{p}_f$ and polarization $\mathbf{e}_f$.
Suppose the initial state of the electrons is $\vert \phi_i \rangle$ at time $t_i$ and
the final state after the scattering is $\vert \phi_f \rangle$ at time $t_f$.
The scattering probability of this two-photon process is described by
\begin{equation}
\Gamma(\mathbf{p}_f \mathbf{e}_f; \mathbf{p}_i \mathbf{e}_i) = \sum_{\phi_i\phi_f}
\frac{1}{Z}e^{-\beta E_i} \left\vert \langle\Psi_F\vert \hat{S}(t_f, t_i) \vert\Psi_I  \rangle \right\vert^{2} ,
\label{eqn1.1}
\end{equation}
where $\hat{S}(t_f, t_i) $ is the time evolution matrix from an initial state
$\vert\Psi_{I} \rangle \equiv \vert\mathbf{p}_i \mathbf{e}_i \phi_i\rangle$
into a final state $ \vert\Psi_{F} \rangle \equiv \vert\mathbf{p}_f \mathbf{e}_f \phi_f\rangle$,
and $E_i$ is the energy of the electrons in the initial state.
Suppose there are $N$ photons in the initial state $\vert\mathbf{p}_i \mathbf{e}_i\rangle$.
Among the $N$ photons there are
$N\sum_{\mathbf{p}_f \mathbf{e}_f} \Gamma(\mathbf{p}_f \mathbf{e}_f; \mathbf{p}_i \mathbf{e}_i) $
photons scattered. The conservation of the photons in the scattering process shows that
\begin{equation}
\Phi_i (\mathbf{p}_i, \mathbf{e}_i) \sigma \Delta t = N \sum_{\mathbf{p}_f \mathbf{e}_f}
\Gamma(\mathbf{p}_f \mathbf{e}_f; \mathbf{p}_i \mathbf{e}_i) ,
\label{eqn1.2}
\end{equation}
where $\sigma$ is the effective scattering cross section,
$\Phi_i (\mathbf{p}_i, \mathbf{e}_i) = nc = \frac{N c}{V} $ is the current density (or flux)
of the incident photons ($V$ is volume of the photon field and $c$ is the light velocity),
and $\Delta t = t_f - t_i$ .
Since $\omega_f = p_f c$, we have $\sum_{\mathbf{p}_f} = \frac{V}{(2\pi c)^{3}}\int \omega_f^{2} d\omega_f d\Omega$
where $d\Omega$  the differential solid angle. The double differential cross
section with the initial and final photon states $\vert\mathbf{p}_i \mathbf{e}_i\rangle$ and
$\vert\mathbf{p}_f \mathbf{e}_f\rangle$ is given by
\begin{equation}
\frac{d^{2}\sigma}{d\Omega d\omega_f}
\Big|_{\mathbf{q}, \nu}
= \frac{V^{2} \omega_f^{2}}{(2\pi)^{3} c^{4}\Delta t}
\Gamma(\mathbf{p}_f \mathbf{e}_f; \mathbf{p}_i \mathbf{e}_i) ,
\label{eqn1.3}
\end{equation}
where $\mathbf{q}$ and $\nu$ are the transferred momentum and energy frequency, respectively,
and are defined by
\begin{equation}
\mathbf{q} = \mathbf{p}_i -\mathbf{p}_{f}, \nu = \omega_i - \omega_f. \label{eqn1.3-2}
\end{equation}
Formula (\ref{eqn1.3}) shows that the differential cross section is proportional to the
scattering probability $\Gamma$. The time difference $\Delta t$ can be canceled
by an additional factor $\Delta t$ in $\Gamma$ which comes from the energy conservation law.
Therefore $\Gamma/\Delta t$ can be taken as a scattering rate.

Suppose the coupling of the electron and the photon field is $V_I$. Define
the total Hamiltonian of the combined system as $\mathscr{H} = H + H_p + V_I $ with
$H$ and $H_p$ the Hamiltonian of the electron and the photon system respectively,
$\hat{S}$ matrix is given by
\begin{equation}
\hat{S}(t_f, t_i) = T_t e^{-\frac{i}{\hbar} \int_{t_i}^{t_f} dt V_I (t) } ,
\label{eqn1.4}
\end{equation}
where $V_I(t) = e^{\frac{i}{\hbar} (H + H_p) (t-t_i) } V_I e^{-\frac{i}{\hbar} (H + H_p)(t-t_i) }$
and $T_t$ is the time ordering operator.
Separate the interaction $V_I$ into $V_1$ of linear to $\mathbf{A}$
and $V_2$ of quadratic to $\mathbf{A}$ , where $\mathbf{A}$ is the photon vector potential.
To lowest-order perturbations, only the following
two expansions of the $\hat{S}$ matrix contribute to the scattering probability $\Gamma$,
\begin{equation}
\Gamma = \sum_{\phi_i\phi_f} \frac{e^{-\beta E_i}}{Z}
\left\vert \langle\Psi_F\vert \hat{S}_1 + \hat{S}_2 \vert\Psi_I  \rangle \right\vert^{2} ,
\label{eqn1.5}
\end{equation}
where $\hat{S}_{1,2}$ are defined by
\begin{eqnarray}
\hat{S}_1 & = & -\frac{i}{\hbar} \int_{t_i}^{t_f} dt  V_{2}(t) , \label{eqn1.6}   \\
\hat{S}_2 & = & \frac{1}{2!}\left(-\frac{i}{\hbar}\right)^{2} \int_{t_i}^{t_f} dt_1 dt_2
T_t \left[ V_{1}(t_1) V_{1}(t_2) \right] . \nonumber
\end{eqnarray}
Thus the scattering probability $\Gamma$ involves  three
contributions,
\begin{equation}
\Gamma = \Gamma_1 + \Gamma_2 + \Gamma_{12} , \label{eqn1.7}
\end{equation}
with
\begin{eqnarray}
\Gamma_1 &=& \sum_{\phi_i\phi_f} \frac{e^{-\beta E_i}}{Z}
\left\vert \langle\Psi_F\vert \hat{S}_1  \vert\Psi_I  \rangle \right\vert^{2} ,  \nonumber \\
\Gamma_2 &=&\sum_{\phi_i\phi_f} \frac{e^{-\beta E_i}}{Z}
\left\vert \langle\Psi_F\vert \hat{S}_2 \vert\Psi_I  \rangle \right\vert^{2} , \label{eqn1.8} \\
\Gamma_{12} &=& \sum_{\phi_i\phi_f} \frac{e^{-\beta E_i}}{Z}
2 \text{Re} \left[ \langle\Psi_I\vert \hat{S}_1^{\dag} \vert\Psi_F  \rangle
\langle\Psi_F\vert \hat{S}_2 \vert\Psi_I  \rangle  \right] . \nonumber
\end{eqnarray}
$\Gamma_1$, $\Gamma_2$ and $\Gamma_{12}$ are the so-called {\it non-resonant},
{\it resonant} and {\it mixed} parts of the scattering probability, respectively.
$\Gamma_{12}$ describes quantum interference of the resonant and non-resonant
scattering processes. The positive or negative $\Gamma_{12}$ comes from the
corresponding constructive or destructive quantum interference.

Since the states of the incident and the scattered photons are defined definitely,
the scattering probability $\Gamma$ can be reduced into a representation
of the pure electron system. Now $\hat{S}_{1,2}$ matrices can be re-expressed in similar
forms to Eq. (\ref{eqn1.6}) where the interactions $V_{1,2}$ are substituted
by the reduced ones $\mathcal{V}_{1,2}$ without photon field involved
(details and derivation will be shown in the following section).

The non-resonant scattering probability $\Gamma_1$ is determined by
the correlation function as
\begin{equation}
\Gamma_1 =  \int_{t_i}^{t_f} d t_1 d t_2
 \langle  \mathcal{V}^{\dag}_{ 2}(t_1) \mathcal{V}_{ 2} (t_2) \rangle , \label{eqn1.9}
\end{equation}
where $\langle \widehat{A} \rangle \equiv \frac{1}{Z} \text{Tr}\left[e^{-\beta H} \widehat{A} \right]$
and $\mathcal{V}(t) = e^{\frac{i}{\hbar} H (t-t_i) } \mathcal{V} e^{-\frac{i}{\hbar} H (t-t_i) }$.
With the fluctuation-dissipation theorem, $\Gamma_1$ can be re-expressed
into the standard form:
\begin{equation}
\Gamma_1 =  \frac{2 \Delta t}{1-e^{-\beta \nu}} \text{Im} \chi(\nu) , \label{eqn1.10}
\end{equation}
where $\chi(\nu)$ is the frequency Fourier transformation of the time-ordered correlation function
$\chi (t_1, t_2) = i \theta (t_1 - t_2) \langle [\mathcal{V}^{\dag}_{2}(t_1),
\mathcal{V}_{2} (t_2) ] \rangle$.
Perturbation theory can then be easily introduced to evaluate $\Gamma_1$.
This is a standard formalism to study the scattering probability in the single-particle
scattering technique such as ARPES, neutron scattering etc.

Because of the time ordering operator $T_t [\mathcal{V}_{1}(t_1) \mathcal{V}_{1} (t_2) ]$
in $\hat{S}_2$ matrix, the fluctuation-dissipation theorem is invalid to evaluate
the resonant $\Gamma_2$ and the mixed $\Gamma_{12}$.
In most studies of the Raman or the X-ray scattering spectrum,
$\Gamma_2$ and $\Gamma_{12}$ are evaluated from the Kramers-Heisenberg formula\cite{DevereauxRMP2007,AmentRMP2011},
where the perturbation is badly controlled and numerical methods are applied.
No reliable perturbation formalism is established even for the weakly interacting electron system.
The time ordering in $\hat{S}_2$ matrix is the difficulty we should overcome to establish
a perturbation formalism to evaluate $\Gamma_2$ and $\Gamma_{12}$.

\begin{figure}[ht]
\includegraphics[width=0.3\columnwidth]{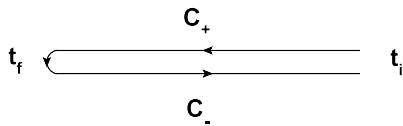}
\caption{ Two-branch contour $C$ for time ordering operator $T_c$. $t_i$ and $t_f$ are the
initial and final times respectively. $C=C_{+}\cup C_{-}$ with an upper time branch
$C_{+}: t_i \rightarrow t_f$ and a lower time branch $C_{-}: t_f \rightarrow t_i$.
If $t_i \rightarrow -\infty, t_f \rightarrow +\infty$, the
contour $C$ is the so-called Schwinger-Keldysh contour\cite{Jorgen}. }
\label{fig1.2}
\end{figure}

From the picture of a time evolution, the scattering probability $\Gamma$ involves two time evolution
processes, forward time ordering from the initial state  $\vert \Psi_i \rangle$ at $t_i$ to
the final state $\vert \Psi_F \rangle$ at $t_f$ , and backward anti-time ordering
from the final state $\vert \Psi_F \rangle$  back to the initial state
$\vert \Psi_I \rangle$. Introducing an anti-time ordering evolution matrix $\widetilde{S}$,
\begin{equation}
\widetilde{S}(t_i, t_f) = \widetilde{T}_{t} e^{-\frac{i}{\hbar} \int_{t_f}^{t_i} dt V_I (t) } ,
\label{eqn1.11}
\end{equation}
we can then define the anti-time ordering $\widetilde{S}_{1,2}$ matrices analog to
$\hat{S}_{1,2}$ in Eq. (\ref{eqn1.6}).
Following the time-and-anti-time evolution picture, $\Gamma_2$ can be expressed as
\begin{eqnarray}
\Gamma_2 &=& \sum_{\phi_i\phi_f} \frac{e^{-\beta E_i}}{Z}
\langle\Psi_I\vert \widetilde{S}_2 \vert\Psi_F  \rangle
\langle\Psi_F\vert \hat{S}_2 \vert\Psi_I  \rangle \nonumber \\
&=& \frac{(-i)^4}{2!^{2}} \int_{[t_i t^{\prime}_i]}
\langle  \widetilde{T}_t [\mathcal{V}^{\dag}_{1}(t^{\prime}_2)\mathcal{V}^{\dag}_{1}(t^{\prime}_1) ]
 T_t [\mathcal{V}_{1}(t_1)\mathcal{V}_{1}(t_2) ]  \rangle ,  \nonumber
\end{eqnarray}
where $\int_{[t_i t^{\prime}_i]} \equiv \int_{t_f}^{t_i} d t^{\prime}_1
d t^{\prime}_2 \int_{t_i}^{t_f} d t_1 d t_2$
(this abbreviation will be used in the whole article where the time variable without or with prime
will follow time evolution or anti-time evolution, respectively).
Introduce a time contour $C$ which describes both the time and the anti-time evolution process,
$C=C_{+}\cup C_{-}$, where $t\in C_{+}$ evolve as $t_i\rightarrow t_f$ and $t^{\prime}\in
C_{-}$ evolve as $t_f\rightarrow t_i$ as shown schematically in Fig. \ref{fig1.2}.
Then $\Gamma_2$ can be re-expressed as
\begin{eqnarray}
\Gamma_2 = \frac{1}{4} \int_{[t_i t^{\prime}_i]}
\langle  T_c [\mathcal{V}_{1}^{\dag}(t^{\prime}_2)\mathcal{V}_{1}^{\dag}(t^{\prime}_1)
\mathcal{V}_{1}(t_1)\mathcal{V}_{1}(t_2) ]  \rangle . \label{eqn1.12}
\end{eqnarray}
Here $T_c$ is the contour time ordering operator defined by
\begin{equation}
T_c [A(t_1) B(t_2)] = \left\{
\begin{array} {l l l}
A(t_1) B(t_2) , &  \text{if} & t_1 >_c t_2 , \\
\pm B(t_2) A(t_1) , &  \text{if} & t_1 <_c t_2 ,
\end{array}
\right.
\label{eqn1.13}
\end{equation}
where $>_c$ and $<_c$ are defined according to the position of the contour time arguments,
latter or earlier in the time contour $C$, and $\pm$ are defined for the bosonic or
fermionic operator, respectively.

From a similar derivation, the non-resonant $\Gamma_1$ and the mixed $\Gamma_{12}$
can be re-expressed within the contour-time formalism as
\begin{eqnarray}
\Gamma_1 &=& - \int_{[t t^{\prime}]}
\langle  T_c [\mathcal{V}_{2}^{\dag}(t^{\prime}) \mathcal{V}_{2}(t) ]  \rangle , \label{eqn1.13-2} \\
\Gamma_{12} &=&  \text{Re} \left [ i  \int_{[t_i t^{\prime}]}
\langle  T_c [\mathcal{V}_{2}^{\dag}(t^{\prime}) \mathcal{V}_{1}(t_1)\mathcal{V}_{1}(t_2) ]\rangle \right] .
\label{eqn1.14}
\end{eqnarray}

Introducing the contour time formalism to evaluate the scattering probability $\Gamma$
is our principle to overcome the difficulty in theoretical study of the two-photon
inelastic light scattering. Formulae (\ref{eqn1.3}), (\ref{eqn1.12}), (\ref{eqn1.13-2})
and (\ref{eqn1.14}) constitute the contour time formalism. When the times are set as
$t_i \rightarrow -\infty$ and $t_f\rightarrow +\infty$, this formalism becomes
a Schwinger-Keldysh formalism which we will use in realistic calculation.

Introduction of the Schwinger-Keldysh contour time formalism
into the study of the resonant inelastic X-ray scattering can
retrospect to 1974 by Nozi$\grave{e}$res and Abrahams\cite{AbrahamsPRB1974},
which was then followed by Igarashi {\it et al} in 2006\cite{IgarashiPRB2006}.
In their formalism the authors focus on the scattering rate,
a time derivative of $\Gamma$ with respective to $t_f$. This scattering rate as
a time derivative leads to the broken equivalence of the time variables and thus
the formalism obtained is limited in study for its complex.
In our formalism, the double differential scattering cross section is
related directly to the scattering probability $\Gamma$ and thus all
the time variables are in equivalent symmetry.
A path integral functional formalism is provided recently by H. C. Lee\cite{LeeNJP2013,Lee2015}.
While only the resonant scattering process is studied for the resonant inelastic light scattering measurement\cite{LeeNJP2013}, all the two-photon scattering processes including the resonant,
non-resonant and mixed responses from magnons in antiferromagnetic insulators are considered
uniformly\cite{Lee2015}. In the article, we present an equivalent canonical formalism to
study the electronic Raman scattering, where all of the electronic responses, the non-resonant
charge density response, the elastic Rayleigh scattering, the fluorescence, the intrinsic
energy-shift Raman scattering and the mixed response, are included in our theory.

The article is arranged as following. In Section \ref{sec1}, we present the principle to
establish a Schwinger-Keldysh contour time formalism for the two-photon inelastic light scattering.
In Section \ref{secsc}, we show an example how to use this formalism to study the electronic
Raman scattering off an one-band superconductor. The scattering cross section with contribution
from all the two-photon processes is studied in details in mean-field approximation.
Summary is present in Section \ref{sect}.
In \ref{appd1}, we provide a preliminary introduction to the non-equilibrium
contour time formalism\cite{Jorgen} for those who are not familiar with it.

\section{Electronic Raman scattering off an one-band superconductor} \label{secsc}

In this section, we show how to study the electronic Raman scattering off
an one-band superconductor with the Schwinger-Keldysh formalism we have
established in Section \ref{sec1}. The contribution from the resonant,
non-resonant and mixed responses to the scattering cross section is
evaluated in details. All the responses are considered in mean-field
approximation in superconducting state as an example.

It should be noted that in superconducting state some low-energy
excitations may play roles in the electronic Raman responses, which include
the Bardasis-Schrieffer bound states\cite{Bardasis, Klein, Monien, Devereaux},
the longitudinal and transverse phase modes\cite{Klein, Monien, Devereaux},
the amplitude Higgs mode\cite{Littlewood} as well as the orbital
excitations in multi-orbital superconductors\cite{Blumberg}.
The Bardasis-Schrieffer bound states are difficult to be resolved in experiments
due to the small binding energy, the finite life time and/or the weak spectrum weight.
The longitudinal phase mode which is important for
the gauge invariance is modified into  high-energy plasma by the Coulomb
interaction and becomes low-energy irrelevant in Raman responses.
Since the main features of the Raman responses are dominated by the gapped
Cooper pairs in superconducting state with a transverse renormalization\cite{Klein, Monien, Devereaux},
in this article we make a mean-field approximation with only pairing interaction
involved. This mean-field approximation is also suitable for simplicity to show how to
calculate the Raman responses with the Schwinger-Keldysh formalism we have established.
The roles of the relevant low-energy excitations in the Raman responses are
other important issues to be studied in future.

\subsection{Scattering probability in contour time formalism} \label{secsc1}

Consider an one-band electron system with Hamiltonian $H=H_t + H_I$, where
$H_t$ and $H_I$ are the free kinetic and the interaction part of the Hamiltonian
respectively. $H_t$ is given by
\begin{equation}
H_t = -\sum_{ij\sigma} t_{ij} d^{\dag}_{i\sigma} d_{j\sigma} , \label{sc1.1}
\end{equation}
where $d_{i\sigma}, d^{\dag}_{i\sigma}$ are the annihilation and creation operators,
respectively, of the electron at site $i$ with spin $\sigma$.
The electron-photon coupling can be obtained by considering the gauge
invariance of $H_t$, which leads to an additional phase factor for $t_{ij}$ and
thus
\begin{equation}
H_t(\mathbf{A}) = -\sum_{ij\sigma} t_{ij}e^{i\frac{e}{\hbar}\mathbf{A}_{ij}\cdot (\mathbf{R}_{j}-\mathbf{R}_i)}
 d^{\dag}_{i\sigma} d_{j\sigma} , \label{sc1.2}
\end{equation}
where $\mathbf{A}_{ij}=\mathbf{A} \left(\frac{\mathbf{R}_{i}+\mathbf{R}_j}{2}\right)$
is defined on bond and the charge of electron is $-e$.
Extend the phase factor into second-order of $\mathbf{A}$,
the electron-photon coupling $V_I = V_1 + V_2$ can be obtained as
\begin{eqnarray}
V_1 &=& \sum_{\mathbf{q}\alpha} j^{\alpha}(-\mathbf{q}) A^{\alpha}(\mathbf{q}) , \label{sc1.3} \\
V_2 &=& \sum_{\mathbf{q}_i \alpha\beta} n^{\alpha\beta}(-\mathbf{q}_1-\mathbf{q}_2)
A^{\alpha}(\mathbf{q}_1) A^{\beta}(\mathbf{q}_2) , \nonumber
\end{eqnarray}
where $\alpha,\beta = x,y,z$.  $j^{\alpha}(-\mathbf{q})$  and $n^{\alpha\beta}(-\mathbf{q}) $
are the current and the stress tensor operator which couple linearly or quadratically to $\mathbf{A} $
respectively, and are defined by
\begin{eqnarray}
j^{\alpha}(-\mathbf{q}) &=& \frac{1}{\sqrt{N}}\sum_{\mathbf{k}\sigma} v^{\alpha}(\mathbf{k,q})
d^{\dag}_{\mathbf{k+q}\sigma} d_{\mathbf{k}\sigma} , \label{sc1.4-0}  \\
n^{\alpha\beta}(-\mathbf{q}) &=& \frac{1}{N} \sum_{\mathbf{k}\sigma}
T^{\alpha\beta}(\mathbf{k,q}) d^{\dag}_{\mathbf{k+q}\sigma} d_{\mathbf{k}\sigma} , \nonumber
\end{eqnarray}
with
\begin{eqnarray}
v^{\alpha}(\mathbf{k,q}) &=& \sum_{\boldsymbol{\delta}} i\frac{e}{\hbar} t_{i, i+ \boldsymbol{\delta}}
\boldsymbol{\delta}^{\alpha} e^{i \left(\mathbf{k} + \frac{\mathbf{q}}{2}\right)\cdot \boldsymbol{\delta}} , \label{sc1.4} \\
T^{\alpha\beta}(\mathbf{k,q}) &=& \sum_{\boldsymbol{\delta}} \frac{1}{2!}\left(\frac{e}{\hbar}\right)^{2}
t_{i, i+ \boldsymbol{\delta}}\boldsymbol{\delta}^{\alpha} \boldsymbol{\delta}^{\beta}
e^{i \left(\mathbf{k} + \frac{\mathbf{q}}{2}\right)\cdot \boldsymbol{\delta}} . \notag
\end{eqnarray}
Note that the electron-photon coupling in (\ref{sc1.3}) only involves the electron charge
degree of freedom. In a more general case, the electron-photon coupling should also
involve the electron magnetic orbital and spin degrees of freedom\cite{AmentRMP2011}.
A similar derivation can be done for the multi-orbital electron
system such as Fe-based superconductors, where the orbital fluctuations may have
unusual contribution to the Raman scattering.

Let us now derive the scattering probability.
Introduce the second quantization of the vector potential $\mathbf{A}$ as\cite{Bruus}
$$
\mathbf{A}(\mathbf{q},t) = \sum_{\lambda}\sqrt{\frac{\hbar}{2\epsilon_0 \omega_{\mathbf{q} \lambda} V}}
\mathbf{e}_{\lambda}(\mathbf{q}) \left(a_{\mathbf{q}\lambda}(t) + a^{\dag}_{\mathbf{q}\lambda}(t) \right) ,
$$
where $a_{\mathbf{q}\lambda}(t) = a_{\mathbf{q}\lambda}e^{-i\omega_{\mathbf{q}\lambda}t}$
and $\mathbf{e}_{\lambda}(\mathbf{q})$ is the polarization vector with $\lambda=1,2$
(the polarization can be linear or circular). In the below
we will assume that the photon energy is polarization independent, i.e.,
$\omega_{\mathbf{q}\lambda} = \omega_{\mathbf{q}}$.
When the photon states in Eq. (\ref{eqn1.6})$\sim $(\ref{eqn1.8}) are traced out,
the non-resonant scattering probability $\Gamma_1$ follows as
\begin{equation}
\Gamma_1 = \sum_{\phi_i\phi_f} \frac{e^{-\beta E_i}}{Z} \left\vert (-i)\int_{t_i}^{t_f} dt
\langle \phi_f \vert \mathcal{V}_2(t) \vert \phi_i \rangle  \right\vert^2 ,\label{sc1.5}
\end{equation}
where the operator $\mathcal{V}_2(t)$ is defined by
\begin{equation}
\mathcal{V}_2(t) =\frac{1}{N} \sum_{\mathbf{k}\sigma} \Lambda(\mathbf{k,q}) T_0(t)
d^{\dag}_{\mathbf{k+q}\sigma} d_{\mathbf{k}\sigma} . \label{sc1.6}
\end{equation}
Here $T_0(t)$ and $\Lambda(\mathbf{k,q})$ are given by
\begin{eqnarray}
T_0(t)  \equiv  e^{-i\nu t}, \Lambda(\mathbf{k,q})  \equiv  \frac{1}{2\epsilon_0 V \sqrt{\omega_i \omega_f}}
\sum_{\alpha\beta} T^{\alpha\beta} (\mathbf{k,q}) \left( e^{*\alpha}_f e^{\beta}_i
+e^{*\beta}_f e^{\alpha}_i \right) ,    \notag
\end{eqnarray}
and $\mathbf{q},\nu$ are the transferred momentum and frequency defined in (\ref{eqn1.3-2}).
Following the principle to establish the contour time formalism in Section \ref{sec1}
and in the non-transient approximation\cite{Jorgen} (shown in \ref{secappd1.1}), $\Gamma_1$
can be expressed in contour time formalism as
\begin{equation}
\Gamma_1 = - \int_{\left[ t^{\prime} t\right]} \langle T_c \hat{S}_c
\mathcal{V}_2^{\dag}(t^{\prime}) \mathcal{V}_2 (t) \rangle _0 , \label{sc1.8}
\end{equation}
where $T_c$ is the time ordering operator defined in the Schwinger-Keldysh contour
$C$ and $t\in C_{+}, t^{\prime}\in C_{-}$.
The contour time evolution operator $\hat{S}_c$ is defined in contour $C$ as
\begin{equation}
\hat{S}_c = T_c e^{-\frac{i}{\hbar}\int_c dt H_I (t)} . \label{sc1.9}
\end{equation}
In formula (\ref{sc1.8}), the operators are defined in interaction representation
by $H_0$ and $\langle A \rangle_0\equiv \frac{\text{Tr}\left[e^{-\beta H_0}A\right]}{Z_0}$.

The resonant scattering probability $\Gamma_2$ is shown to follow
$$
\Gamma_2 = \sum_{\phi_i\phi_f} \frac{e^{-\beta E_i}}{Z} \left\vert
\frac{(-i)^2}{2!} \sum_l \int_{\left[t_1 t_2\right]}
\langle \phi_f \vert \pi_l(t_1, t_2) \vert \phi_i \rangle  \right\vert^2 ,
$$
where $\pi_l(t_1, t_2)$ is given by
\begin{equation}
\pi_l(t_1, t_2)  = T_t \left[ \mathcal{V}_{l1} (t_1) \mathcal{V}_{l2}(t_2) \right] . \label{sc1.9-2}
\end{equation}
The $l$-dependent interactions $\mathcal{V}_{l1}$ and $\mathcal{V}_{l2}$ ($l=1,2$) are
defined by
\begin{eqnarray}
\mathcal{V}_{l j} (t) = \sum_{\lambda\alpha} \frac{1}{\sqrt{2\epsilon_0 V \hbar \omega_{q_j}}}
e^{\alpha}_{\lambda}(\mathbf{q}_j) e^{-i\omega_j t} j^{\alpha} (-\mathbf{q}_j, t) . \label{sc1.9-3}
\end{eqnarray}
Here $\mathbf{q}_j,\omega_j, j=1,2$ are all $l$-dependent defined in Table \ref{tabsc1}.
In the Schwinger-Keldysh contour time formalism, $\Gamma_2$ can be expressed in the
non-transient approximation as
\begin{equation}
\Gamma_2 = \frac{1}{4} \sum_{ll^{\prime}}\int_{\left[ t_i^{\prime} t_i \right]}
\langle T_c \hat{S}_c \pi_{l^{\prime}}^{\dag}(t_1^{\prime}, t_2^{\prime})
\pi_l (t_1, t_2) \rangle _0 , \label{sc1.10}
\end{equation}
where $t_1, t_2\in C_{+}, t_1^{\prime}, t_2^{\prime}\in C_{-}$.
Following a similar procedure, the mixed scattering probability $\Gamma_{12}$
can be expressed in the contour time formalism as
\begin{eqnarray}
\Gamma_{12} = \text{Re}\left[ i \sum_{l}\int_{\left[ t^{\prime} t_i \right]}
\langle T_c \hat{S}_c \mathcal{V}_{2}^{\dag}(t^{\prime}) \pi_l (t_1, t_2) \rangle _0 \right] , \label{sc1.12}
\end{eqnarray}
with $t_1, t_2\in C_{+}, t^{\prime}\in C_{-}$.

\begin{table}[ht]
\caption{ Parameters $\mathbf{q}_j,\omega_j, j=1,2$  of $\mathcal{V}_{l j}(t)$  in $l$-dependent.
Also included are the $l$-dependent time factor $T_l(t_1, t_2)$. }
\label{tabsc1}
\begin{ruledtabular}
\begin{tabular}{cccccc}
$l$  &  $\mathbf{q}_1$ & $\mathbf{q}_2$ & $\omega_1$ & $\omega_2$  & $T_l(t_1, t_2)$ \\
\hline
1 & $-\mathbf{p}_f$ & $\mathbf{p}_i$ & $-\omega_f$ & $\omega_i$ & $e^{i\omega_f t_1 - i\omega_i t_2}$  \\
2 & $\mathbf{p}_i$ & $-\mathbf{p}_f$ & $\omega_i$ & $-\omega_f$ & $e^{i\omega_f t_2 - i\omega_i t_1} $
\end{tabular}
\end{ruledtabular}
\end{table}

\subsection{ Green's functions in superconducting state } \label{secsc2}

In this article, we focus our study on the Raman responses of a superconductor in
superconducting state. Thus we can simplify the electron Hamiltonian $H=H_t + H_I$
in mean-field approximation as
\begin{equation}
H_0 = \sum_{\mathbf{k}} \Psi_{\mathbf{k}}^{\dag} \left(\varepsilon_{\mathbf{k}} \tau_3
+ \Delta_{\mathbf{k}} \tau_1 \right)\Psi_{\mathbf{k}} , \label{sc2.1}
\end{equation}
where $\Psi_{\mathbf{k}}$ is the so-called Nambu spinor defined as
$\Psi_{\mathbf{k}} = \left( d_{\mathbf{k}\uparrow}, d^{\dag}_{-\mathbf{k}\downarrow} \right)^{T}$,
$\varepsilon(\mathbf{k})$ is the band energy, and $\Delta_{\mathbf{k}}$
is the paring potential of the Cooper pairs. $\tau_i, i=1,2,3$ are the Pauli matrices.
Introduce transformation $\Psi_{\mathbf{k}} = U(\mathbf{k}) \Phi_{\mathbf{k}}$ where
$\Phi_{\mathbf{k}} = \left( f_{\mathbf{k}\uparrow}, f_{-\mathbf{k}\downarrow} \right)^{T}$,
$H_0$ can be diagonalized into the form as
\begin{equation}
H_0 = \sum_{\mathbf{k}} \Phi_{\mathbf{k}}^{\dag} \left[ E_{\mathbf{k}} \tau_3 \right] \Phi_{\mathbf{k}} ,
\label{sc2.2}
\end{equation}
where $E_{\mathbf{k}} = \sqrt{\varepsilon^{2}_{\mathbf{k}} + \Delta^{2}_{\mathbf{k}}}$ is the
diagonalized energy. Transformation matrix is defined as
$U(\mathbf{k}) = u_{\mathbf{k}} - i\tau_3 v_{\mathbf{k}},$
with the matrix elements
$u_{\mathbf{k}} = \sqrt{\frac{1}{2}\left( 1 +\frac{\varepsilon_{\mathbf{k}}}{E_{\mathbf{k}}}  \right)}$
and $v_{\mathbf{k}} =\text{sgn}\left( \Delta_{\mathbf{k}} \right)   \sqrt{\frac{1}{2}\left( 1
- \frac{\varepsilon_{\mathbf{k}}}{E_{\mathbf{k}}} \right)}$.

Define the contour time Green's function for the two-component operator
$\Phi_{\mathbf{k}}$ as
\begin{equation}
G_{c} (\mathbf{k};t_1,t_2) = -i \langle T_c \Phi_{\mathbf{k}} (t_1)
 \Phi^{\dag}_{\mathbf{k}} (t_2) \rangle .  \label{sc2.4}
\end{equation}
For realistic calculation, we introduce the corresponding real-time Green's function
as
\begin{equation}
G(\mathbf{k};t_1, t_2) = \left( \begin{array}{cc}
G^{T} (\mathbf{k}; t_1, t_2) & G^{<} (\mathbf{k}; t_1, t_2) \\
G^{>} (\mathbf{k}; t_1, t_2) & G^{\widetilde{T}} (\mathbf{k}; t_1, t_2)
\end{array}
\right) , \label{sc2.5}
\end{equation}
where an additional Schwinger-Keldysh index is introduced according
to whether $t_1, t_2$ belong to $C_{+}$ or $C_{-}$\cite{Jorgen}.
The four real-time Green's functions in $G(\mathbf{k};t_1, t_2)$ are
defined by
\begin{eqnarray}
G^{>} (\mathbf{k}; t_1, t_2) &=& -i \langle \Phi_{\mathbf{k}} (t_1)
 \Phi^{\dag}_{\mathbf{k}} (t_2) \rangle ,  \text{  }
G^{<} (\mathbf{k}; t_1, t_2) = i \langle \Phi^{\dag}_{\mathbf{k}} (t_2)
\Phi_{\mathbf{k}} (t_1) \rangle , \notag \\
G^{T}(\mathbf{k}; t_1, t_2) &=& \theta(t_1-t_2) G^{>}(\mathbf{k}; t_1, t_2)
 + \theta(t_2-t_1)G^{<}(\mathbf{k}; t_1, t_2) , \label{sc2.5-2} \\
G^{\widetilde{T}}(\mathbf{k}; t_1, t_2) &=& \theta(t_2-t_1) G^{>}(\mathbf{k}; t_1, t_2)
+ \theta(t_1-t_2)G^{<}(\mathbf{k}; t_1, t_2) . \notag
\end{eqnarray}
Fourier transformed the real-time Green's function by
$
G(\mathbf{k};t ) = \frac{1}{2\pi} \int_{-\infty}^{\infty} d \omega
G(\mathbf{k},\omega) e^{-i\omega t}  $
and
$
G(\mathbf{k},\omega) = \int_{-\infty}^{\infty} d t
G(\mathbf{k};t) e^{i\omega t }
$
with the real-time translational symmetry, it is readily shown that the real-time
Green's functions in mean-field superconducting state follow
\begin{eqnarray}
&&G_0^{<} (\mathbf{k},\omega) = \left( \begin{array} {cc}
i n_{\mathbf{k}\uparrow} 2\pi \delta(\omega-E_{\mathbf{k}}) & 0 \\
0 & i n_{-\mathbf{k}\downarrow} 2\pi \delta(\omega+E_{\mathbf{k}})
\end{array}
\right) , \notag \\
&& G_0^{>} (\mathbf{k},\omega) = \left( \begin{array} {cc}
-i \left(1-n_{\mathbf{k}\uparrow}\right) 2\pi \delta(\omega-E_{\mathbf{k}}) & 0 \\
0 & -i\left(1- n_{-\mathbf{k}\downarrow}\right) 2\pi \delta(\omega+E_{\mathbf{k}})
\end{array}
\right) , \notag \\
&& G_0^{T} (\mathbf{k},\omega) = \left( \begin{array} {cc}
\frac{1-n_{\mathbf{k}\uparrow}}{\omega-E_{\mathbf{k}}+i\delta^{+}}
+ \frac{n_{\mathbf{k}\uparrow}}{\omega-E_{\mathbf{k}}-i\delta^{+}} & 0 \\
0 & \frac{1-n_{-\mathbf{k}\downarrow}}{\omega+E_{\mathbf{k}}+i\delta^{+}}
+ \frac{n_{-\mathbf{k}\downarrow}}{\omega+E_{\mathbf{k}}-i\delta^{+}}
\end{array}
\right) ,  \label{sc2.6} \\
&& G_0^{\widetilde{T}} (\mathbf{k},\omega) = \left( \begin{array} {cc}
\frac{-\left(1-n_{\mathbf{k}\uparrow}\right)}{\omega-E_{\mathbf{k}}-i\delta^{+}}
+ \frac{-n_{\mathbf{k}\uparrow}}{\omega-E_{\mathbf{k}}+i\delta^{+}} & 0 \\
0 & \frac{-\left(1-n_{-\mathbf{k}\downarrow}\right)}{\omega+E_{\mathbf{k}}-i\delta^{+}}
+ \frac{-n_{-\mathbf{k}\downarrow}}{\omega+E_{\mathbf{k}}+i\delta^{+}}
\end{array}
\right) , \notag
\end{eqnarray}
where $n_{\mathbf{k}\uparrow} = \langle f_{\mathbf{k}\uparrow}^{\dag} f_{\mathbf{k}\uparrow} \rangle_0
=\frac{1}{e^{\beta E_{\mathbf{k}}}+1}$
and $n_{-\mathbf{k}\downarrow} = \langle f_{-\mathbf{k}\downarrow}^{\dag} f_{-\mathbf{k}\downarrow} \rangle_0
=\frac{1}{e^{-\beta E_{\mathbf{k}}}+1}$, and $\delta^{+}$ is a positive infinitesimal value.

\begin{figure}[ht]
\includegraphics[width=0.4\columnwidth]{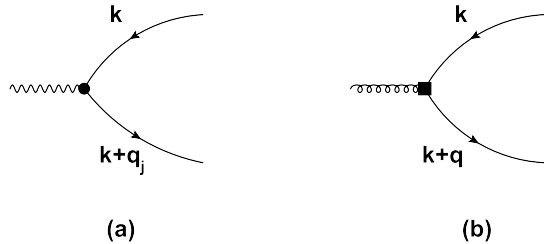}
\caption{ Feynman diagrams for the interaction vertices. (a) for $\mathcal{V}_{l j}$ with
vertex factor $\Upsilon_{1}(\mathbf{k},\mathbf{q}_j)$ and
(b) for $\mathcal{V}_2$ with vertex factor $\Upsilon_{2}(\mathbf{k,q})$.  }
\label{figsc2.1}
\end{figure}

The reduced interactions $\mathcal{V}_{lj}(t)$ and $\mathcal{V}_2(t)$ defined in Section \ref{secsc1}
can now be re-expressed by the new two-component operator $\Phi_{\mathbf{k}}$ as
\begin{eqnarray}
\mathcal{V}_{l j}(t) &=& \frac{1}{\sqrt{N}}\sum_{\mathbf{k}}  \Phi^{\dag}_{\mathbf{k}+\mathbf{q}_j}(t)
\Upsilon_{1} (\mathbf{k},\mathbf{q}_j) \Phi_{\mathbf{k}}(t) e^{-i\omega_j t} , \notag \\
\mathcal{V}_{2}(t)  &=& \frac{1}{N}\sum_{\mathbf{k}}  \Phi^{\dag}_{\mathbf{k+q}}(t)
\Upsilon_{2} (\mathbf{k,q }) \Phi_{\mathbf{k}}(t) e^{-i\nu t}, \label{sc2.7}
\end{eqnarray}
where $\mathbf{q}, \nu$ are the transferred momentum and frequency defined in Eq. (\ref{eqn1.3-2})
and $\mathbf{q}_j, \omega_j (j = 1, 2)$ are $l$-dependent as given in Table \ref{tabsc1}.
$\Upsilon_{1}$ and $\Upsilon_{2}$ are defined by
\begin{eqnarray}
\Upsilon_{1}(\mathbf{k},\mathbf{q}_j) &= & \frac{1}{\sqrt{2\epsilon_0 V \hbar \omega_{q_j}}}\sum_{\lambda\alpha}
e^{\alpha}_{\lambda}(\mathbf{q}_j) v^{\alpha} (\mathbf{k},\mathbf{q}_j), \label{sc2.9}  \\
\Upsilon_{2}(\mathbf{k},\mathbf{q} ) &= &\Lambda(\mathbf{k},\mathbf{q}) \left[ \frac{\varepsilon_{\mathbf{k}}}{E_{\mathbf{k}}} \tau_3 - \frac{\Delta_{\mathbf{k}}}{E_{\mathbf{k}}} \tau_1\right] . \notag
\end{eqnarray}
To obtain $\Upsilon_{2}(\mathbf{k},\mathbf{q} )$, we have considered the approximation
$\mathbf{q}\rightarrow 0$ in Raman scattering.
Fig. \ref{figsc2.1} shows the schematic Feynman diagrams for the interaction vertices.

\subsection{ Non-resonant scattering probability $\Gamma_1$} \label{secsc3}

\begin{figure}[ht]
\includegraphics[width=0.3\columnwidth]{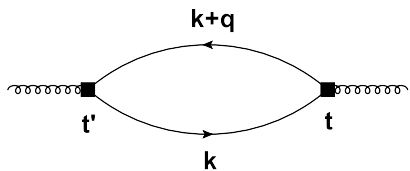}
\caption{ Non-resonant $\Gamma_1$ in mean-field superconducting state.
The solid line with arrow represents the Green's functions (\ref{sc2.6}) and  the
vertex square represents the factor $\Upsilon_2(\mathbf{k,q})$ in $\mathcal{V}_2(t)$. }
\label{figsc3.1}
\end{figure}

Consider the non-resonant scattering probability $\Gamma_1$ with the contour
time formula (\ref{sc1.8}).
In superconducting state, we consider a mean-field approximation
where we neglect the roles of the collective low-energy excitations
as we have discussed above.
$\Gamma_1$ can be approximated in zero-th order mean-filed perturbation as
\begin{equation}
\Gamma_1^{(0)} = - \int_{\left[ t^{\prime} t\right]} \langle T_c
\mathcal{V}_2^{\dag}(t^{\prime}) \mathcal{V}_2 (t) \rangle _0 , \label{sc3.2}
\end{equation}
which is shown schematically by the Feynman diagram in Fig. \ref{figsc3.1}.

Use Wick's theorem to decompose the multi-particle correlation in $\Gamma_1^{(0)}$ and
then transform it into the real-time formalism, the non-resonant scattering probability
$\Gamma_1^{(0)}$ in mean-field approximation is given by
\begin{eqnarray}
&&\Gamma_1^{(0)} = \frac{1}{N^2}\sum_{\mathbf{k}}\int_{t_i}^{t_f} d t d t^{\prime}
\text{Tr} [\Upsilon_2^{*}(\mathbf{k,q}) G_{0}^{>}(\mathbf{k+q};t^{\prime}-t) \notag \\
&&\Upsilon_2(\mathbf{k,q}) G_{0}^{<}(\mathbf{k};t-t^{\prime}) ]
 e^{-i\nu (t-t^{\prime})} \notag \\
&& = \frac{\Delta t}{2\pi N^2} \sum_{\mathbf{k}}\int d\omega
 \text{Tr}[
 \Upsilon_2^{*}(\mathbf{k,q}) G_{0}^{>}(\mathbf{k+q};\omega+\nu)
 \Upsilon_2(\mathbf{k,q}) G_{0}^{<}(\mathbf{k_1};\omega) ] . \label{sc3.3}
\end{eqnarray}
To obtain $\Delta t$ in the last derivation, we have used the identity
$\frac{\Delta t}{2\pi}\delta(\nu - \nu^{\prime}) =
\frac{1}{(2\pi)^2} \int_{t_i}^{t_f} d t d t^{\prime} e^{-i(\nu-\nu^{\prime})(t-t^{\prime})}$.
Substitute the Green's functions Eq. (\ref{sc2.6}) into Eq. (\ref{sc3.3}),
$\Gamma_1^{(0)}$ is shown to be
\begin{eqnarray}
\Gamma_1^{(0)} = \frac{2\pi \Delta t}{N^2}\sum_{\mathbf{k}} \left\vert \Lambda (\mathbf{k,q}) \right\vert^{2}
\left\{ \begin{array}{c}
c_{+-} \delta(\nu+E_{\mathbf{k}}-E_{\mathbf{k+q}})  \\
+c_{-+} \delta(\nu-E_{\mathbf{k}}+E_{\mathbf{k+q}})   \\
+c_{++} \delta(\nu+E_{\mathbf{k}}+E_{\mathbf{k+q}})  \\
+c_{--} \delta(\nu-E_{\mathbf{k}}-E_{\mathbf{k+q}})
\end{array}
\right\} , \label{sc3.4}
\end{eqnarray}
where $c_{\pm\pm}$ are defined by
\begin{eqnarray}
&& c_{+-} = \frac{\varepsilon^{2}_{\mathbf{k}}}{E^{2}_{\mathbf{k}}}
\left( 1-n_{\mathbf{k+q}\uparrow} \right) n_{\mathbf{k}\uparrow} ,
c_{-+} = \frac{\varepsilon^{2}_{\mathbf{k}}}{E^{2}_{\mathbf{k}}}
\left( 1-n_{\mathbf{-k-q}\downarrow} \right) n_{-\mathbf{k}\downarrow} , \notag \\
&&c_{++} = \frac{\Delta^{2}_{\mathbf{k}}}{E^{2}_{\mathbf{k}}}
\left( 1-n_{\mathbf{-k-q}\downarrow} \right) n_{\mathbf{k}\uparrow} ,
c_{--} = \frac{\Delta^{2}_{\mathbf{k}}}{E^{2}_{\mathbf{k}}}
\left( 1-n_{\mathbf{k+q}\uparrow} \right) n_{-\mathbf{k}\downarrow} . \notag
\end{eqnarray}
In formula (\ref{sc3.4}), $c_{+-}$ and $c_{-+}$ terms describe contribution from
single-particle excitations and $c_{++}$ and $c_{--}$ terms  from Cooper pairs.
At low temperature $T\ll T_c$, since $n_{\mathbf{k}\uparrow} = 0$ and
$n_{-\mathbf{k}\downarrow} = 1$, $c_{+-} = c_{-+} = c_{++} = 0$ and $c_{--}$=1.
In this case, only Cooper pairs provide finite contribution to the non-resonant scattering,
and thus
\begin{equation}
\Gamma_1^{(0)} = \frac{2\pi \Delta t}{N^2}\sum_{\mathbf{k}} \left\vert \Lambda (\mathbf{k,q}) \right\vert^{2}
\frac{\Delta^{2}_{\mathbf{k}}}{E^{2}_{\mathbf{k}}} \delta(\nu-E_{\mathbf{k}}-E_{\mathbf{k+q}}) .
\label{sc3.5}
\end{equation}
It shows that there is a threshold frequency $\nu_c$ beyond which
the non-resonant scattering probability is finite. In a $s$-wave superconductor
$\nu_c = 2\Delta$ with $\Delta$ the superconducting gap.
Our result Eq. (\ref{sc3.5}) is same to the previous one in the non-interacting
limit and $\mathbf{q}\rightarrow 0$\cite{Klein}.

\subsection{Resonant scattering probability $\Gamma_2$ } \label{secsc4}

In the mean-field superconducting state, $\Gamma_2$ can be
approximated at zero-th order as
\begin{equation}
\Gamma_2^{(0)} = \frac{1}{4} \sum_{ll^{\prime}}\int_{\left[ t_i^{\prime} t_i \right]}
\langle T_c \pi_{l^{\prime}}^{\dag}(t_1^{\prime}, t_2^{\prime})
\pi_l (t_1, t_2) \rangle _0 . \label{sc4.1}
\end{equation}
Substitute Eq. (\ref{sc1.9-2}) of $\pi_l(t_1, t_2)$ into this formula, $\Gamma_2^{(0)}$ is re-expressed as
\begin{equation}
\Gamma_2^{(0)} = \frac{1}{4N^2}\sum_{\mathbf{k}_j \mathbf{k}^{\prime}_{j} l l^{\prime}}
O_{l^\prime}^{*} (\left\{ \mathbf{k}^{\prime}_{j},\mathbf{q}^{\prime}_{j} \right\})O_{l}(\left\{ \mathbf{k}_j,\mathbf{q}_j \right\})
\widetilde{\Gamma}_{2}^{(l l^{\prime})} , \label{sc4.2}
\end{equation}
where $\mathbf{q}_j, \mathbf{q}^{\prime}_{j},j=1,2$ are $l$-dependent given in Table \ref{tabsc1} and
$O_{l}(\left\{ \mathbf{k}_j,\mathbf{q}_j \right\})\equiv O_{l}(\mathbf{k}_1,\mathbf{q}_1;\mathbf{k}_2,\mathbf{q}_2) $
are defined by
\begin{eqnarray}
O_{l}(\left\{ \mathbf{k}_j,\mathbf{q}_j \right\})
= \Upsilon_{1}(\mathbf{k}_1,\mathbf{q}_1) \Upsilon_{1}(\mathbf{k}_2,\mathbf{q}_2) . \label{sc4.3}
\end{eqnarray}
In formula (\ref{sc4.2}), $\widetilde{\Gamma}_{2}^{(l l^{\prime})}$ is defined by
\begin{eqnarray}
\widetilde{\Gamma}_{2}^{(l l^{\prime})} = &&
\int_{\left[t^{\prime}_{j} t_j\right]} T^{*}_{l^{\prime}} (t^{\prime}_1,t^{\prime}_2) T_{l}(t_1, t_2)
\langle T_c   \Phi_{\mathbf{k}^{\prime}_2}^{\dag} (t^{\prime}_2)
\Phi_{\mathbf{k}^{\prime}_2 + \mathbf{q}^{\prime}_2} (t^{\prime}_2) \Phi_{\mathbf{k}^{\prime}_1}^{\dag} (t^{\prime}_1)
\Phi_{\mathbf{k}^{\prime}_1 + \mathbf{q}^{\prime}_1} (t^{\prime}_1) \notag \\
&&\Phi_{\mathbf{k}_1 + \mathbf{q}_1}^{\dag} (t_1)  \Phi_{\mathbf{k}_1}(t_1)
\Phi_{\mathbf{k}_2 + \mathbf{q}_2}^{\dag} (t_2)  \Phi_{\mathbf{k}_2}(t_2) \rangle_0  , \label{sc4.4}
\end{eqnarray}
where $t_1, t_2\in C_{+}$, $t^{\prime}_1,t^{\prime}_2 \in C_{-}$ and the $l$-dependent variables $T_{l} (t_1, t_2) = e^{-i\omega_1 t_1 - i\omega_2 t_2}$ are given in Table \ref{tabsc1}.

In the following, we will make all the Wick's decompositions for the many-particle correlation
in $\widetilde{\Gamma}_{2}^{(l l^{\prime})}$. They can be classified into three categories,
the Rayleigh scattering, the fluorescence and the intrinsic energy-shift resonant Raman scattering.

\subsubsection{Rayleigh scattering} \label{secsc4.1}

\begin{figure}[ht]
\includegraphics[width=0.25\columnwidth]{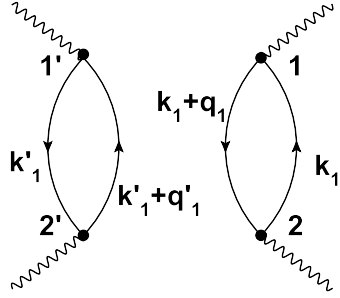}
\caption{ Feynman diagram for elastic Rayleigh scattering.  $1, 2$ ($1^{\prime}, 2^{\prime}$)
on the vertices represent the time $t_1$, $t_2$ ($t_1^{\prime}$, $t_2^{\prime}$)
in the forward (backward) time contour $C_{+}$ ($C_{-}$). }
\label{figsc4.1b}
\end{figure}

Rayleigh scattering is an elastic scattering with the incident and scattered photons
having same frequency. The Feynman diagram for Rayleigh scattering is Fig. \ref{figsc4.1b},
which shows us that
\begin{eqnarray}
\widetilde{\Gamma}^{(l l^{\prime})}_{2,1} = && \frac{\Delta t}{2\pi} \int_{\nu^{\prime}_1 \nu_1}
\text{Tr} \left[ G_0^{\widetilde{T}}(\mathbf{k}^{\prime}_1,\nu^{\prime}_1)
G_0^{\widetilde{T}}(\mathbf{k}^{\prime}_1+ \mathbf{q}^{\prime}_1,\nu^{\prime}_1+\omega^{\prime}_1)  \right] \notag\\
&&\text{Tr} \left[ G_0^{T}(\mathbf{k}_1,\nu_1)
G_0^{T}(\mathbf{k}_1+ \mathbf{q}_1,\nu_1+\omega_1) \right] \delta_r , \label{sc4.5}
\end{eqnarray}
where $\int_{\nu^{\prime}_1 \nu_1} = \int_{-\infty}^{+\infty} d \nu^{\prime}_1 d \nu_1$
and $\delta_r = \delta_{\mathbf{k}_2+\mathbf{q}_2, \mathbf{k}_1}  \delta_{\mathbf{q}_2, -\mathbf{q}_1}
\delta_{\mathbf{k}^{\prime}_2+\mathbf{q}^{\prime}_2, \mathbf{k}^{\prime}_1}
\delta_{\mathbf{q}^{\prime}_2, -\mathbf{q}^{\prime}_1}$ .

The Rayleigh scattering probability denoted by $\Gamma_{2,1}^{(0)}$ is shown to be
\begin{equation}
\Gamma_{2,1}^{(0)} = \frac{2\Delta t}{\pi} \left\vert I_1 \right\vert^2, \label{sc4.6}
\end{equation}
where the exchange symmetry between $l (l^{\prime}) = 1, 2$ has been considered and $I_1$ is defined by
\begin{equation}
I_1 = \frac{1}{N}\sum_{\mathbf{k}} \int d\omega
O_{1} \text{Tr} \left[ G_0^{T}(\mathbf{k}, \omega)
G_0^{T}(\mathbf{k}- \mathbf{p}_i,\omega-\omega_i) \right]. \label{sc4.7}
\end{equation}
Here $\mathbf{p}_i$ and $\omega_i$ are the momentum and frequency of the incident photons
and $O_{1}\equiv O_{1}(\mathbf{k},-\mathbf{p}_i;\mathbf{k}-\mathbf{p}_i,\mathbf{p}_i) $.
In the superconducting state, $I_1$ follows
\begin{eqnarray}
I_1 = \frac{2\pi i}{N}\sum_{\mathbf{k}} O_{1}
\left\{
\begin{array}{c}
\frac{(1-n_{\mathbf{k}\uparrow}) n_{\mathbf{k}-\mathbf{p}_i \uparrow}}
{\omega_i + E_{\mathbf{k}-\mathbf{p}_i}-E_{\mathbf{k}}+i\delta^{+}}
- \frac{n_{\mathbf{k}\uparrow}(1-n_{\mathbf{k}-\mathbf{p}_i \uparrow})}
{\omega_i + E_{\mathbf{k}-\mathbf{p}_i}-E_{\mathbf{k}}-i\delta^{+}} \notag\\
+\frac{(1-n_{-\mathbf{k}\downarrow}) n_{-\mathbf{k}+\mathbf{p}_i \downarrow}}
{\omega_i - E_{\mathbf{k}-\mathbf{p}_i}+E_{\mathbf{k}}+i\delta^{+}}
- \frac{n_{-\mathbf{k}\downarrow}(1-n_{-\mathbf{k}+\mathbf{p}_i \downarrow})}
{\omega_i - E_{\mathbf{k}-\mathbf{p}_i}+E_{\mathbf{k}}-i\delta^{+}}
\end{array}
\right\} . \label{sc4.8}
\end{eqnarray}
This formula shows clearly that only the single-particle excitations have dominant
contribution to the elastic Rayleigh scattering in superconducting
state. At low temperature $T \ll T_c$, the finite superconducting gap leads to
$I_1 = 0$ and thus the Rayleigh scattering is strongly suppressed.

It should be noted that since $I_1 = O(N^0)=O(1)$, the Rayleigh scattering probability
$\Gamma_{2,1}^{(0)}$ is also in order of $O(1)$. This is in contrast to the
non-resonant scattering probability $\Gamma_1^{(0)}$ in Eq. (\ref{sc3.4}) or (\ref{sc3.5}) which
is in order of $O(1/N)$. If there is no special mechanism to suppress the
Rayleigh scattering, it will be several orders of magnitude larger than the
non-resonant scattering in contribution to the scattering cross section.

\subsubsection{Fluorescence} \label{secsc4.2}

\begin{figure}[ht]
\includegraphics[width=0.25\columnwidth]{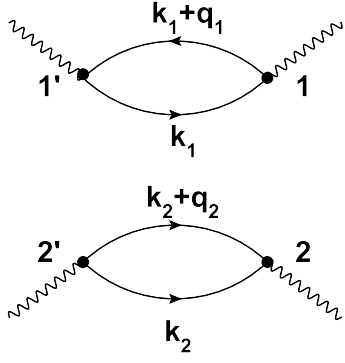}
\caption{ Feynman diagram for fluorescence scattering. }
\label{figsc4.2}
\end{figure}

In a fluorescence process, there are sequent photon absorption and photon emission
as schematically shown by the Feynman diagram in Fig. \ref{figsc4.2}.
Wick's decomposition for the fluorescence scattering shows that
\begin{eqnarray}
\widetilde{\Gamma}^{(l l^{\prime})}_{2,2} = && 2\times \frac{\Delta t}{2\pi} \int_{\nu_1 \nu_2}
\text{Tr} \left[ G_0^{>}(\mathbf{k}_1+\mathbf{q}_1,\nu_1+\omega_1) G_0^{<}(\mathbf{k}_1,\nu_1)  \right] \notag\\
&&\text{Tr} \left[ G_0^{>}(\mathbf{k}_2+\mathbf{q}_2,\nu_2+\omega_2) G_0^{<}(\mathbf{k}_2,\nu_2)  \right]
\delta_f , \label{sc4.9}
\end{eqnarray}
where
$
\delta_f = \delta_{\mathbf{k}_1,\mathbf{k}^{\prime}_1} \delta_{\mathbf{k}_2,\mathbf{k}^{\prime}_2}
\delta_{\mathbf{q}_1,\mathbf{q}^{\prime}_1} \delta_{\mathbf{q}_2,\mathbf{q}^{\prime}_2} \delta_{l,l^{\prime}}
$
and $\mathbf{q}_j,\omega_j$ are $l$-dependent as given in Table \ref{tabsc1}.

The fluorescence scattering probability denoted by $\Gamma_{2,2}^{(0)}$ is shown to be
\begin{equation}
\Gamma_{2,2}^{(0)} = \frac{\Delta t}{2\pi} I^{(2)}_{1,1} I^{(2)}_{1,2} ,  \label{sc4.10}
\end{equation}
where $l$-dependent integral $I^{(2)}_{l,j}$ is defined by
\begin{equation}
I^{(2)}_{l,j} = \frac{1}{N}\sum_{\mathbf{k}} \int d\nu_1 \left\vert \Upsilon_1(\mathbf{k},\mathbf{q}_j) \right\vert^2
\text{Tr} \left[ G_0^{>}(\mathbf{k}+\mathbf{q}_j,\nu_1+\omega_j) G_0^{<}(\mathbf{k},\nu_1)  \right] .
\label{sc4.11}
\end{equation}
In the mean-field superconducting state, $I^{(2)}_{l,j}$ follows
\begin{equation}
I^{(2)}_{l,j} = \frac{(2\pi)^2}{N}\sum_{\mathbf{k}} \left\vert \Upsilon_1(\mathbf{k},\mathbf{q}_j) \right\vert^2
\left\{ \begin{array}{c}
n_{\mathbf{k}\uparrow}(1-n_{\mathbf{k}+\mathbf{q}_j \uparrow})
\delta(\omega_j + E_{\mathbf{k}+\mathbf{q}_j}-E_{\mathbf{k}}) \\
 + n_{-\mathbf{k}\downarrow}(1-n_{-\mathbf{k}-\mathbf{q}_j \downarrow})
\delta(\omega_j - E_{\mathbf{k}+\mathbf{q}_j}+E_{\mathbf{k}})
\end{array}
\right\} . \label{sc4.12}
\end{equation}
It shows that only the single-particle excitations have contribution to the fluorescence
scattering. At low temperature $T\ll T_c$, $I^{(2)}_{l,j}=0$, thus the fluorescence
response is largely suppressed in superconducting state. Moreover, the fluorescence
scattering probability  $\Gamma_{2,2}^{(0)} $ is in order of $O(1)$ similar to
the Rayleigh scattering $\Gamma_{2,1}^{(0)}$.

\subsubsection{Intrinsic resonant Raman scattering} \label{secsc4.3}


\begin{figure}[ht]
\includegraphics[width=0.5\columnwidth]{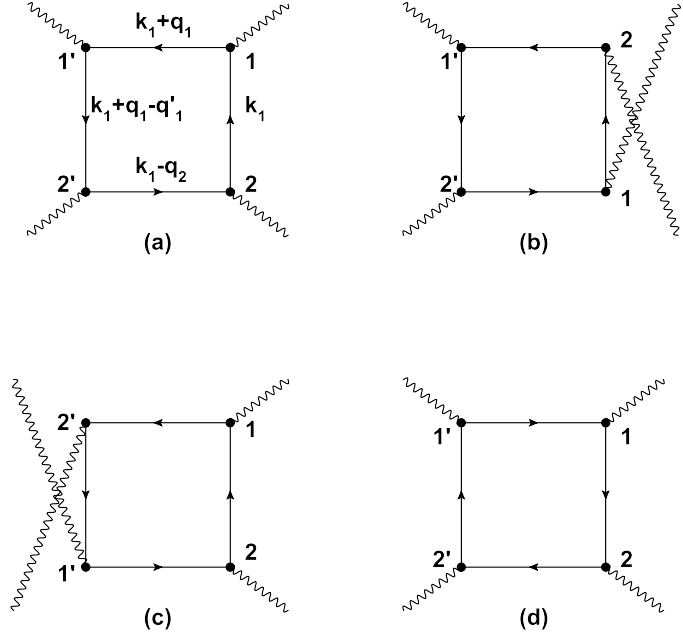}
\caption{ Feynman diagrams for intrinsic resonant Raman scattering without time cross. }
\label{figsc4.3}
\end{figure}

\begin{figure}[ht]
\includegraphics[width=0.5\columnwidth]{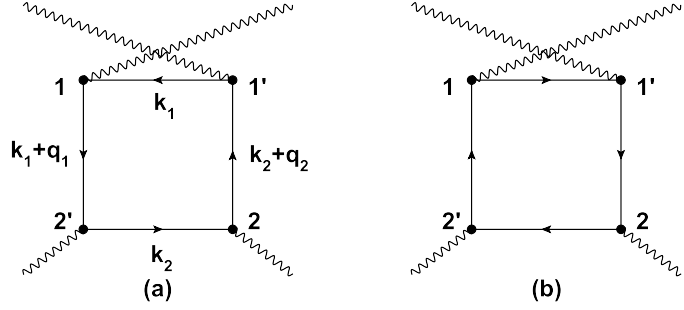}
\caption{ Feynman diagrams for intrinsic resonant Raman scattering with time cross. }
\label{figsc4.5}
\end{figure}

The intrinsic resonant Raman scattering processes are shown schematically in
Fig. \ref{figsc4.3} and \ref{figsc4.5}. They are all one-loop Feynman diagrams with
four vertices of $\mathcal{V}_{lj}$ and are classified into the two categories,
ones shown in Fig. \ref{figsc4.3} where the times in contour branch $C_{+}$
do not cross the times in $C_{-}$ and the others shown in Fig. \ref{figsc4.5}
with times cross.

Denote the intrinsic resonant Raman scattering without time evolution cross by
$\Gamma_{2,3}^{(0)}$. The four Wick's decompositions for $\Gamma_{2,3}^{(0)}$
as shown schematically  by the four Feynman diagrams in Fig. \ref{figsc4.3}
have same contribution because the exchange of $l (l^{\prime}) = 1, 2 $
is equivalent to the exchange of the time arguments. Therefore, we need only
to consider one Feynman diagram such as Fig. \ref{figsc4.3} (a)
with an additional factor $4$. The Raman scattering probability
$\Gamma_{2,3}^{(0)}$ is shown to be
\begin{equation}
\Gamma_{2,3}^{(0)} = \frac{1}{4N^2}\sum_{\mathbf{k}_1 l l^{\prime}}
O_{l^\prime}^{*} O_{l} \widetilde{\Gamma}_{2,3}^{(l l^{\prime})} , \label{sc4.14}
\end{equation}
where $O_{l^\prime}^{*} = O_{l^\prime}^{*} (\mathbf{k}_1 + \mathbf{q}_1 -\mathbf{q}_1^{\prime},
\mathbf{q}^{\prime}_1; \mathbf{k}_1 - \mathbf{q}_2,
\mathbf{p}_i - \mathbf{p}_f -\mathbf{q}_1^{\prime})$,
$O_{l}=O_{l}(\mathbf{k}_1,\mathbf{q}_1 ; \mathbf{k}_1 - \mathbf{q}_2,\mathbf{q}_2)$,
$\left\{\mathbf{q}_{j},\omega_j, \mathbf{q}_{j}^{\prime},\omega_j^{\prime}\right\}, j=1,2$
are $l$-dependent given in Table \ref{tabsc1}.
$\widetilde{\Gamma}^{(l l^{\prime})}_{2,3}$ is given by
\begin{eqnarray}
\widetilde{\Gamma}^{(l l^{\prime})}_{2,3} = && - 4\times\frac{\Delta t}{2\pi} \int_{\nu_j^{\prime} \nu_j}
\text{Tr} [ G_0^{\widetilde{T}}(\mathbf{k}_1+\mathbf{q}_1-\mathbf{q}^{\prime}_{1},
\nu_2^{\prime}) G_0^{>}(\mathbf{k}_1 + \mathbf{q}_1,\nu_1^{\prime}) \notag \\
&&G_0^{T}(\mathbf{k}_1,\nu_1) G_0^{<}(\mathbf{k}_1-\mathbf{q}_2,\nu_2)   ]
 \delta_{r\nu} , \label{sc4.15}
\end{eqnarray}
where $\int_{\nu_j^{\prime} \nu_j} = \int_{-\infty}^{+\infty} d\nu_1^{\prime} \nu_2^{\prime} d\nu_1  d\nu_2$
and $\delta_{r\nu} = \delta(\nu_2^{\prime}-\nu_1 -\omega_1 + \omega_1^{\prime})
\delta(\nu^{\prime}_1-\nu_1 -\omega_1) \delta(\nu_2 -\nu_1 + \omega_2)$.

In the mean-field superconducting state, $\widetilde{\Gamma}^{(l l^{\prime})}_{2,3}$ follows
\begin{eqnarray}
\widetilde{\Gamma}^{(l l^{\prime})}_{2,3}& = & 8\pi \Delta t [ I_{3,1}\delta(\omega_f-\omega_i
+E_{\mathbf{k}_1+\mathbf{q}_1}-E_{\mathbf{k}_1-\mathbf{q}_2}) \notag \\
&& + I_{3,2} \delta(\omega_f-\omega_i
-E_{\mathbf{k}_1+\mathbf{q}_1}+E_{\mathbf{k}_1-\mathbf{q}_2}) ], \label{sc4.16}
\end{eqnarray}
where $I_{3,j}$ are defined by
\begin{eqnarray}
 I_{3,1} &=&  \left(\frac{1-n_{\mathbf{k}_1 + \mathbf{q}_1 -\mathbf{q}_1^{\prime} \uparrow}}
{\omega_1^{\prime} - E_{\mathbf{k}_1+\mathbf{q}_1}+ E_{\mathbf{k}_1 + \mathbf{q}_1 -\mathbf{q}_1^{\prime}}+i\delta^{+}}
+  \frac{n_{\mathbf{k}_1 + \mathbf{q}_1 -\mathbf{q}_1^{\prime}\uparrow}}
{\omega_1^{\prime} - E_{\mathbf{k}_1+\mathbf{q}_1}+E_{\mathbf{k}_1 + \mathbf{q}_1 -\mathbf{q}_1^{\prime}}-i\delta^{+}}  \right) \notag \\
&& \times \left(\frac{1-n_{\mathbf{k}_1\uparrow}}
{\omega_1 - E_{\mathbf{k}_1+\mathbf{q}_1}+E_{\mathbf{k}_1}-i\delta^{+}}
+  \frac{n_{\mathbf{k}_1 \uparrow}}
{\omega_1 - E_{\mathbf{k}_1+\mathbf{q}_1}+E_{\mathbf{k}_1}+i\delta^{+} }  \right) \notag \\
&&\left( 1-n_{\mathbf{k}_1 + \mathbf{q}_1 \uparrow} \right) n_{\mathbf{k}_1 - \mathbf{q}_2 \uparrow} , \notag \\
I_{3,2} &=&  \left(\frac{1-n_{- \mathbf{k}_1 - \mathbf{q}_1 + \mathbf{q}_1^{\prime} \downarrow}}
{\omega_1^{\prime} + E_{\mathbf{k}_1+\mathbf{q}_1}-E_{\mathbf{k}_1 + \mathbf{q}_1 -\mathbf{q}_1^{\prime}}+i\delta^{+}}
+  \frac{n_{- \mathbf{k}_1 - \mathbf{q}_1 +\mathbf{q}_1^{\prime} \downarrow}}
{\omega_1^{\prime} + E_{\mathbf{k}_1+\mathbf{q}_1}-E_{\mathbf{k}_1 + \mathbf{q}_1 -\mathbf{q}_1^{\prime}}-i\delta^{+}}  \right) \notag \\
&& \times \left(\frac{1-n_{-\mathbf{k}_1\downarrow}}
{\omega_1 + E_{\mathbf{k}_1+\mathbf{q}_1}-E_{\mathbf{k}_1}-i\delta^{+}}
+  \frac{n_{-\mathbf{k}_1 \downarrow}}
{\omega_1 + E_{\mathbf{k}_1+\mathbf{q}_1}-E_{\mathbf{k}_1}+i\delta^{+} }  \right) \notag \\
&&\left( 1-n_{-\mathbf{k}_1 - \mathbf{q}_1 \downarrow} \right) n_{-\mathbf{k}_1 + \mathbf{q}_2 \downarrow} . \notag
\end{eqnarray}
Similar to the Rayleigh and fluorescence scattering processes, only the single-particle excitations
contribute to the non-time-cross intrinsic resonant Raman scattering. At low temperature $T \ll T_c$, the finite
superconducting gap strongly suppresses the scattering probability $\Gamma_{2,3}^{(0)}$.


From a similar derivation, the time-cross intrinsic resonant Raman scattering probability
denoted by $\Gamma_{2,4}^{(0)}$ is shown to follow
\begin{equation}
\Gamma_{2,4}^{(0)} = -\frac{1}{4 N^2}\sum_{\mathbf{k}_1 l l^{\prime}}
O_{l^\prime}^{*} O_{l} \widetilde{\Gamma}_{2,4}^{(l l^{\prime})} , \label{sc4.20}
\end{equation}
where $O_{l^\prime}^{*} = O_{l^\prime}^{*} (\mathbf{k}_1,\mathbf{q}^{\prime}_1; \mathbf{k}_1 - \mathbf{q}_2 + \mathbf{q}_1^{\prime}, \mathbf{p}_i - \mathbf{p}_f - \mathbf{q}_1^{\prime} )$
and $O_{l}(\mathbf{k}_1,\mathbf{q}_1 ; \mathbf{k}_1 - \mathbf{q}_2 + \mathbf{q}_1^{\prime},\mathbf{q}_2)$.
The two Wick's decompositions shown by the two Feynman diagrams in Fig. \ref{figsc4.5}
leads to
\begin{eqnarray}
\widetilde{\Gamma}^{(l l^{\prime})}_{2,4} &=&  2\times \frac{\Delta t}{2\pi} \int_{\nu_j^{\prime} \nu_j}
\text{Tr} [ G_0^{>}(\mathbf{k}_1+\mathbf{q}_1, \nu_1^{\prime})
G_0^{<}(\mathbf{k}_1,\nu_1) \notag \\
&&G_0^{>}(\mathbf{k}_2 + \mathbf{q}_2,\nu_2^{\prime})
G_0^{<}(\mathbf{k}_2,\nu_2) ] \delta_{\mathbf{k}_2, \mathbf{k}_1 - \mathbf{q}_2 + \mathbf{q}_1^{\prime}}
\delta_{c} , \label{sc4.21}
\end{eqnarray}
where $\delta_{c} = \delta(\nu_1^{\prime}-\nu_1 -\omega_1) \delta(\nu^{\prime}_2-\nu_1 -\omega_1^{\prime})
\delta(\nu_2 -\nu_1 -\omega_1 + \omega_2^{\prime})$.

In the mean-field superconducting state, $\widetilde{\Gamma}^{(l l^{\prime})}_{2,4} $ follows
\begin{equation}
\widetilde{\Gamma}^{(l l^{\prime})}_{2,4} = 2\times (2\pi)^3 \Delta t \left\{
\begin{array}{c}
I_{4,1} \delta(\omega_1+E_{\mathbf{k}_1} -E_{\mathbf{k}_1+\mathbf{q}_1})
\delta(\omega_1^{\prime}+ E_{\mathbf{k}_1} -E_{\mathbf{k}_1+\mathbf{q}_1^{\prime}}) \\
\delta(\omega_2 - \omega_1^{\prime} - E_{\mathbf{k}_1} + E_{\mathbf{k}_1 - \mathbf{q}_2 + \mathbf{q}_1^{\prime}}) \\
+ I_{4,2} \delta(\omega_1-E_{\mathbf{k}_1} +E_{\mathbf{k}_1+\mathbf{q}_1})
\delta(\omega_1^{\prime}-E_{\mathbf{k}_1} +E_{\mathbf{k}_1+\mathbf{q}_1^{\prime}}) \\
\delta(\omega_2 - \omega_1^{\prime}+E_{\mathbf{k}_1} -E_{\mathbf{k}_1 - \mathbf{q}_2 + \mathbf{q}_1^{\prime}})
\end{array}
 \right\}, \label{sc4.22}
\end{equation}
where $I_{4,j}$ are defined by
\begin{eqnarray}
&&I_{4,1} = (1-n_{\mathbf{k}_1+\mathbf{q}_1 \uparrow})n_{\mathbf{k}_1\uparrow}
(1-n_{\mathbf{k}_1+\mathbf{q}_1^{\prime} \uparrow})n_{\mathbf{k}_1 - \mathbf{q}_2 + \mathbf{q}_1^{\prime} \uparrow} \notag \\
&&I_{4,2} = (1-n_{-\mathbf{k}_1-\mathbf{q}_1 \downarrow})n_{-\mathbf{k}_1\downarrow}
(1-n_{-\mathbf{k}_1-\mathbf{q}_1^{\prime} \downarrow})n_{-\mathbf{k}_1 + \mathbf{q}_2 - \mathbf{q}_1^{\prime}\downarrow} . \notag
\end{eqnarray}
At low temperature $T\ll T_c$, since $n_{\mathbf{k}\uparrow}=0$ and $n_{-\mathbf{k}\downarrow} = 1$,
$I_{4,j}=0$. Therefore $\Gamma_{2,4}^{(0)}$ is strongly suppressed in superconducting state.
Note that all the contributions from the intrinsic resonant Raman scattering to the cross section are
in order of $O(1/N)$, in contrast to that from the two-loop Rayleigh and fluorescence processes.

In the general effective mass approximation for the electronic Raman scattering\cite{DevereauxRMP2007,Abrikosov1973},
the resonant and non-resonant responses are described by one uniform Raman charge density.
In that approximation, the contributions from the resonant and non-resonant responses
would be in proportion to each other. This is obviously in contrast to our results,
where in superconducting state the non-resonant response has finite contribution from
Cooper pairs while the resonant response has only contribution from single-particle
excitations and is strongly suppressed by the superconducting gap.

\subsection{ Mixed scattering probability $\Gamma_{12}$ } \label{secsc5}

\begin{figure}[ht]
\includegraphics[width=0.5\columnwidth]{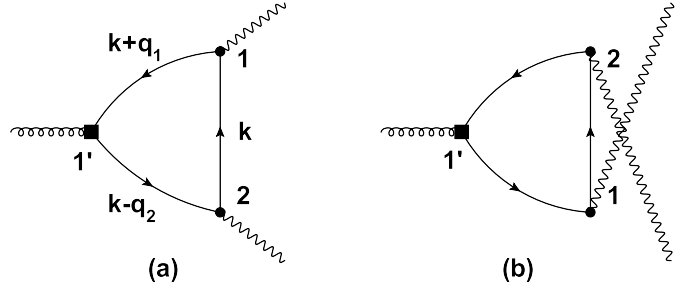}
\caption{ Feynman diagrams for resonant-non-resonant mixed scattering. }
\label{figsc5.1}
\end{figure}

The mixed scattering is a pure quantum effect, as it comes from the quantum interference
of the resonant and non-resonant scattering processes. In the mean-field superconducting state,
the mixed scattering probability denoted by $\Gamma_{12}^{(0)}$ can be approximate by
\begin{eqnarray}
\Gamma_{12}^{(0)} = \text{Re}\left[ i \sum_{l}\int_{\left[ t^{\prime} t_i \right]}
\langle T_c  \mathcal{V}_{2}^{\dag}(t^{\prime}) \pi_l (t_1, t_2) \rangle _0 \right] . \label{sc5.1}
\end{eqnarray}
There are two Wick's decompositions for $\Gamma_{12}^{(0)}$ as shown in Fig. \ref{figsc5.1}.
Because of the equivalence of the exchange of the time arguments
and the exchange of $l = 1, 2$, the two Feynman diagrams have same contribution to the scattering
probability. Thus we only need to consider one Feynman diagram with an additional
factor $2$. $\Gamma_2^{(0)}$ is shown to be
\begin{equation}
\Gamma_{12}^{(0)} = \text{Re}\left[ \frac{1}{N^2}\sum_{\mathbf{k}  l }
\Lambda^{*}(\mathbf{k} - \mathbf{q}_2,\mathbf{p}_i - \mathbf{p}_f)
O_{l}(\mathbf{k},\mathbf{q}_1; \mathbf{k} - \mathbf{q}_2,\mathbf{q}_2 )
\widetilde{\Gamma}_{12}^{(l)} \right] , \label{sc5.6}
\end{equation}
where $\mathbf{q}_j,j=1,2$ are $l$-dependent given in Table \ref{tabsc1}
and $\widetilde{\Gamma}_{12}^{(l)}$ is given by
\begin{eqnarray}
\widetilde{\Gamma}_{12}^{(l)} &=& 2\times \frac{\Delta t}{2\pi} \int d\nu_1
\text{Tr} [
\left(\frac{\varepsilon_{\mathbf{k}^{\prime}}}{E_{\mathbf{k}^{\prime}}} \tau_3
- \frac{\Delta_{\mathbf{k}^{\prime}}}{E_{\mathbf{k}^{\prime}}} \tau_1\right)
G_{0}^{>}(\mathbf{k} + \mathbf{q}_1,\nu_1+\omega_1) G_0^{T} (\mathbf{k},\nu_1) \notag \\
&&G_{0}^{<}(\mathbf{k} - \mathbf{q}_2,\nu_1-\omega_2)
] \delta_{\mathbf{k}^{\prime},\mathbf{k}-\mathbf{q}_2} . \label{sc5.4}
\end{eqnarray}

In the mean-field superconducting state, $\widetilde{\Gamma}_{12}^{(l)}$ follows
\begin{eqnarray}
\widetilde{\Gamma}_{12}^{(l)} &=& \left(4\pi \Delta t\right) \frac{ \varepsilon_{\mathbf{k} -\mathbf{q}_2}}
{E_{\mathbf{k} -\mathbf{q}_2}}
[ - I_{5,1} \delta(\omega_f -\omega_i + E_{\mathbf{k}+\mathbf{q}_1} - E_{\mathbf{k}-\mathbf{q}_2} ) \notag \\
&&+I_{5,2} \delta(\omega_f -\omega_i - E_{\mathbf{k}+\mathbf{q}_1} + E_{\mathbf{k}-\mathbf{q}_2} )], \label{sc5.5}
\end{eqnarray}
where $I_{5,j},j=1,2$ are defined by
\begin{eqnarray}
&&I_{5,1} = (1-n_{\mathbf{k}+\mathbf{q}_1 \uparrow})  n_{\mathbf{k}-\mathbf{q}_2 \uparrow}
\left( \begin{array}{c}
\frac{1-n_{\mathbf{k}\uparrow}}{\omega_1 - E_{\mathbf{k}+\mathbf{q}_1} + E_{\mathbf{k}} - i\delta^{+}} \\
+ \frac{n_{\mathbf{k}\uparrow}}{\omega_1 - E_{\mathbf{k}+\mathbf{q}_1} + E_{\mathbf{k}} + i\delta^{+}}
\end{array}
\right) , \notag \\
&&I_{5,2} = (1-n_{-\mathbf{k}-\mathbf{q}_1 \downarrow})  n_{-\mathbf{k}+\mathbf{q}_2 \downarrow}
\left( \begin{array}{c}
\frac{1-n_{-\mathbf{k}\downarrow}}{\omega_1 + E_{\mathbf{k}+\mathbf{q}_1} - E_{\mathbf{k}} - i\delta^{+}} \\
+ \frac{n_{-\mathbf{k}\downarrow}}{\omega_1 + E_{\mathbf{k}+\mathbf{q}_1} - E_{\mathbf{k}} + i\delta^{+}}
\end{array}
\right) . \notag
\end{eqnarray}
It shows that only the single-particle excitations have contribution to the mixed scattering
probability in superconducting state. At low temperature $T \ll T_c$, $I_{5,j}=0$, thus
the mixed scattering probability $\Gamma_{12}^{(0)}$ is strongly suppressed.
It should be noted that $\Gamma_{12}^{(0)}$ may be positive or negative in
accord with the constructive or destructive interference. Moreover, it has
magnitude in order of $O(1/N)$ similar to the resonant scattering with one-loop Feynman diagram.

\section{Summary} \label{sect}

In the above sections, we have present a Schwinger-Keldysh perturbation formalism for
the electronic Raman scattering. All the two-photon scattering processes can be well
included within this contour time formalism and the contributions from the resonant,
non-resonant and mixed responses can be studied uniformly. As an example, we evaluate the
Raman scattering cross section off an one-band superconductor. In the mean-field
superconducting state, Cooper pairs contribute only to the non-resonant response.
All the other responses from the Rayleigh  scattering, the fluorescence, the intrinsic
energy-shift resonant Raman scattering and the mixed scattering are dominated by
the single-particle excitations and are strongly suppressed by the superconducting gap.
The above formalism can be easily extended for the high-energy X-ray scattering
when the inner core electrons are included to couple with the photon field\cite{AmentRMP2011}.
A similar Schwinger-Keldysh formalism can then be established with a similar
procedure.




\appendix

\section{Review of Schwinger-Keldysh contour time formalism } \label{appd1}

In this Appendix, we review the Schwinger-Keldysh contour time formalism which has been
well established for non-equilibrium physics. This is a preliminary introduction for those
who are not familiar with this formalism. More details can be found in Rammer's textbook\cite{Jorgen}.

\subsection{Contour time formalism} \label{secappd1.1}

Our task is to calculate the contour time correlation function defined by
\begin{equation}
O_c = \langle T_c [A_H(t_1) B_H (t_2) \cdots C_H (t_3)] \rangle , \label{appd1.0}
\end{equation}
where $A_H (t) = e^{\frac{i}{\hbar} H (t-t_i)} A e^{-\frac{i}{\hbar} H (t-t_i)}$.
The Hamiltonian of the system is defined by
\begin{equation}
H = H_0 + H_I , \label{appd1.1}
\end{equation}
where $H_0$ is the quadratic part which can be treated exactly and $H_I$ includes all
the left such as the scattering potential and the inter-particle
interaction, etc.

\begin{figure}[ht]
\includegraphics[width=0.3\columnwidth]{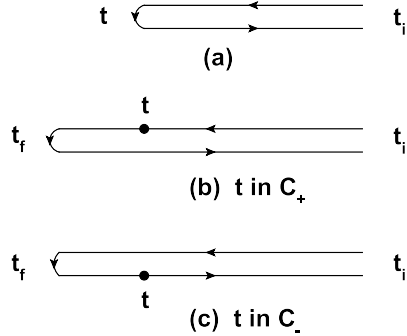}
\caption{ Schematic illustration of the contour time representation of
$ A_H (t)$. (a) for $A_H (t)=\hat{S}(t_i,t) A_{H_0}(t) \hat{S}(t, t_i)$,
(b) and (c) for $ A_H (t) = T_c[\hat{S}_c A_{H_0}(t) ] $ with $t \in C_{+}$ and $t\in C_{-}$ respectively. }
\label{fig-appd1.1}
\end{figure}

Introduce the time evolution $\hat{U}$ matrix as
$\hat{U}(t, t_i) = e^{\frac{i}{\hbar} H_0(t-t_i)} e^{-\frac{i}{\hbar} H(t-t_i)}$
and $\hat{S}$ matrix as $\hat{S}(t_2, t_1) = \hat{U}(t_2, t_i) \hat{U}^{\dag}(t_1, t_i)$,
then
$$
A_H (t) = \hat{U}^{\dag}(t,t_i) A_{H_0}(t) \hat{U}(t, t_i) =\hat{S}(t_i,t) A_{H_0}(t) \hat{S}(t, t_i),
$$
where $A_{H_0} (t) = e^{\frac{i}{\hbar} H_0 (t-t_i)} A e^{-\frac{i}{\hbar} H_0 (t-t_i)}$.
In the contour time formalism, $A_H (t)$ can be re-expressed  as
\begin{equation}
A_H (t) = T_c[\hat{S}_c A_{H_0}(t) ], \label{appd1.2}
\end{equation}
where $\hat{S}_c$ matrix is defined in the time contour $C$ as
\begin{equation}
\hat{S}_c = T_c e^{-\frac{i}{\hbar} \int_c d t H_I(t) } .  \label{appd1.3}
\end{equation}
Here $H_I(t) = e^{\frac{i}{\hbar} H_0 (t-t_i)} H_I e^{-\frac{i}{\hbar} H_0 (t-t_i)} $ and
the contour time ordering operator $T_c$ is defined in contour $C=C_{+}\cup C_{-}$
with $\int_c d t \equiv \int_{t_i\rightarrow t_f\rightarrow t_i} d t$.
To obtain the contour representation (\ref{appd1.2}), we have use the transitivity of the
$\hat{S}$ matrix $\hat{S}(t_3, t_1) = \hat{S}(t_3, t_2) \hat{S}(t_2, t_1)$. For example, if $t\in C_{+}$,
$A_H (t) = \hat{S}(t_i,t) A_{H_0}(t) \hat{S}(t, t_i) = \hat{S}(t_i,t_f) \hat{S}(t_f,t)
A_{H_0}(t) \hat{S}(t, t_i) = T_c[\hat{S}_c A_{H_0}(t) ]$,
and if $t\in C_{-}$, $A_H (t) = \hat{S}(t_i,t) A_{H_0}(t) \hat{S}(t, t_i) = \hat{S}(t_i,t) A_{H_0}(t)
\hat{S}(t, t_f) \hat{S}(t_f, t_i)  = T_c[\hat{S}_c A_{H_0}(t) ]$.
The contour time representation (\ref{appd1.2}) is illustrated schematically in Fig. \ref{fig-appd1.1}.

Following this principle, the contour time ordered correlation function $O_c$ in (\ref{appd1.0}) can be
expressed as
\begin{eqnarray}
O_c  = \langle T_c [\hat{S}_c A(t_1) B (t_2) \cdots C (t_3)] \rangle , \label{appd1.4}
\end{eqnarray}
where the subscript $H_0$ in the operator $A, B$ and $C$ has been ignored for clarity.

\begin{figure}[ht]
\includegraphics[width=0.3\columnwidth]{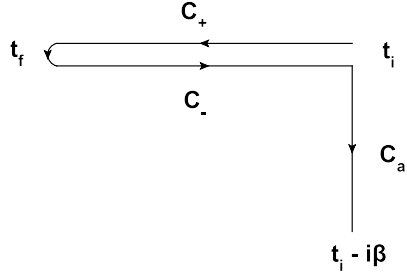}
\caption{ Three-branch contour (Kadanoff-Baym contour) $C_{i}=C_{+}\cup C_{-} \cup C_{a}$,
where $C_{a}$ is the imaginary appendix contour defined as $C_a: t_i\rightarrow t_i - i\beta$. }
\label{fig-appd1.2}
\end{figure}

Now let us consider the thermal average. Since $e^{-\beta H} = e^{-\beta H_0} U_{\tau}$
with $U_{\tau} = e^{\beta H_0} e^{-\beta H} $. Introduce an imaginary time axis,
$U_{\tau}$ can be expressed as
$$ U_{\tau} = \hat{S}_{c_a}  = T_{c_a} e^{-\frac{i}{\hbar} \int_{c_a} dt H_I (t)}, $$
where the appendix contour $C_{a}$ is defined as $C_a: t_i \rightarrow t_i - i\beta$
(shown in Fig. \ref{fig-appd1.2}) and the integral
$\int_{c_a} dt = \int_{t_i\rightarrow t_i - i\beta} dt$.
The correlation function $O_c$ follows
\begin{eqnarray}
O_c &=& \frac{ \text{Tr}\left[ e^{-\beta H_0} \hat{S}_{c_a} T_c [\hat{S}_c A(t_1) B (t_2) \cdots C (t_3)] \right] }
{\text{Tr}\left[ e^{-\beta H_0} \hat{S}_{c_a}\right] }  \nonumber  \\
&=& \frac{ \text{Tr}\left[ e^{-\beta H_0} T_{c_i} [\hat{S}_{c_i} A(t_1) B (t_2) \cdots C (t_3)] \right] }
{\text{Tr}\left[ e^{-\beta H_0} \hat{S}_{c_i}\right] } ,
\label{appd1.5}
\end{eqnarray}
where $T_{c_i}$ is the time ordering operator in the contour $C_i = C_{+}\cup C_{-}\cup C_{a}$
and $\hat{S}_{c_i}$ matrix is defined by
$$\hat{S}_{c_i}  = T_{c_i} e^{-\frac{i}{\hbar} \int_{c_i} dt H_I (t)}.$$
When we are not interested in the transient physics on the collision time scale,
we can set $t_i\rightarrow -\infty$,
then the contribution from the imaginary part of the contour $C_i$ vanishes
due to the thermal fluctuations\cite{Jorgen}. The contour time correlation
function can be approximated into
\begin{eqnarray}
O_c = \langle T_c [\hat{S}_c A(t_1) B (t_2) \cdots C (t_3)] \rangle_0 . \label{appd1.6}
\end{eqnarray}
This is the {\it non-transient} approximation of the contour time correlation function.
The time contour $C$ in our study will be defined as the so-called
Schwinger-Keldysh contour where $t_i\rightarrow -\infty, t_f\rightarrow +\infty$.
$\langle A \rangle_0 $ is defined by
$\langle A \rangle_0 = \frac{1}{Z_0} \text{Tr}\left[e^{-\beta H_0} A\right]$.

When expand the $\hat{S}_c$ matrix in order of $H_I$ and decompose the many-particle
correlation function by Wick's theorem, the perturbation corrections to $O_c$ can be
obtained order-by-order in principle.
Wick's theorem in the contour time formalism has been shown to have
a same manner to the ground-state and the finite-temperature formalism.
Define the single-electron Green's function as
\begin{equation}
G_c(1,2) = -i \langle T_c d_{1} d_{2}^{\dag} \rangle, \label{appd1.7}
\end{equation}
where indices $1,2$ involve the momentum, spin and temporal indices, etc.
The zero-th order Green's function is denoted by $G_0(1,2)$.
Wick's theorem leads to all possible decompositions in an example as below:
\begin{eqnarray}
& & \langle T_c d_1 d_2  d_3^{\dag} d_4^{\dag} \rangle_0  \nonumber \\
&& =\langle T_c  d_1  d_4^{\dag} \rangle_0  \langle T_c  d_2  d_3^{\dag} \rangle_0
   \pm \langle T_c  d_1  d_3^{\dag} \rangle_0  \langle T_c d_2  d_4^{\dag} \rangle_0 \nonumber \\
&&= - G_0(1,4) G_0(2,3) \mp G_0(1,3) G_0(2,4) . \label{appd1.8}
\end{eqnarray}
where $\mp$ in last equation correspond to the bosonic and fermionic fields respectively.

\subsection{Real-time formalism} \label{secappd1.2}

The above formalism provides principle for the contour time perturbation
theory. In realistic calculation, we will introduce the corresponding real-time
formalism. In this formalism, the single-particle contour time ordered Green's
function $G_c(1,2)$ is transformed into a $2\times 2$ matrix Green's function $G(1,2)$,
\begin{equation}
G(1,2) = \left(
\begin{array} {cc}
G_{11}(1,2) & G_{12}(1,2) \\
G_{21}(1,2) & G_{22}(1,2)
\end{array}
\right) ,
\label{appd1.9}
\end{equation}
where the subscribe indices $n,m$ in $G_{nm}(1,2)$ are the so-called Schwinger-Keldysh
indices and are defined as $n (m) = 1, 2$
according to $t_1 (t_2)\in C_{+} $ or $ C_{-}$.
The real-time Green's function has another familiar denotation
\begin{equation}
G(1,2) = \left(
\begin{array} {cc}
G^{T}(1,2) & G^{<}(1,2) \\
G^{>}(1,2) & G^{\widetilde{T}}(1,2)
\end{array}
\right) ,
\label{appd1.10}
\end{equation}
where the matrix element Green's functions are defined by
\begin{eqnarray}
&&G^{>}(1,2)=  -i \langle  d_{1} d_{2}^{\dag} \rangle ,  \text{ }
G^{<}(1,2)= i \langle  d_{2}^{\dag} d_{1} \rangle , \label{appd1.11} \\
&&G^{T}(1,2)= -i \langle T_t  d_{1}  d_{2}^{\dag}\rangle , \text{ }
G^{\widetilde{T}}(1,2)=  -i \langle \widetilde{T}_t  d_{1}  d_{2}^{\dag}\rangle . \nonumber
\end{eqnarray}
It can be shown easily that
\begin{eqnarray}
&&G^{T}(1,2)= \theta(t_1-t_2) G^{>}(1,2) +  \theta(t_2-t_1) G^{<}(1,2),  \nonumber \\
&&G^{\widetilde{T}}(1,2)= \theta(t_1-t_2) G^{<}(1,2) +  \theta(t_2-t_1) G^{>}(1,2). \nonumber
\end{eqnarray}

The above formalism is defined for the fermionic filed. A similar formalism
can be established for the bosonic field, where Bose-Einstein statistics should be
introduced. Moreover the perturbation expansions in the real-time matrix
formalism can be obtained one-to-one from the expansions in the contour
time formalism.





\end{document}